\documentclass[aps,prb,twocolumn,groupedaddress,eqsecnum,showpacs]{revtex4}

\usepackage{amsfonts}
\usepackage{dcolumn}
\usepackage{graphicx,amssymb,amsmath}
\usepackage{bm}
\usepackage{natbib}
\usepackage{epstopdf}
\begin{document}

\title{ Perturbative criteria for Anderson localization in long-ranged 1D tight-binding models}

\author{Shimul Akhanjee}
\email[]{shimul@physics.ucla.edu}

\affiliation{Department of Physics, UCLA, Box 951547, Los Angeles, CA 90095-1547}


\date{\today}

\begin{abstract}
We develop an alternative scaling approach to determine the criteria for Anderson localization in one-dimensional tight-binding models with random site energies having a bandwidth that decays as a power law in space, $H_{ij} \propto \left|i - j\right|^{-\alpha}$. At the first order in perturbation theory the scale dependence of the exchange-narrowed energy of the disorder is compared to the energy level spacing of the ideal system to establish whether or not the disorder has a perturbative effect on the Bloch states. We find that at $\alpha =1$, the perturbative condition is satisfied and for sufficiently weak disorder strength all states are extended. For $\alpha > 1$, all states are localized for arbitrary disorder strength, in agreement with the earlier renormalization group treatment by Levitov.
 
\end{abstract}


\maketitle

\section{introduction}

The subject of strong localization of electronic states was pioneered by early efforts of P. W. Anderson (1958)\cite{Anderson}, revealing the existence of a phase transition from a metallic phase to an insulating phase and established a remarkable set of criteria or bounds on the transition in terms of the relative disorder strength. In one-dimension (1D), it  can be rigorously shown that there is a strict absence of extended states for an arbitrarily finite amount of disorder. This was first stated by Mott and Twose (1961) in a heuristic manner\cite{motttwose} and later made more rigorous by others\cite{ locreview93}. However, the validity of this result only encompasses models with uncorrelated disorder and nearest-neighbor hopping terms.

Recent efforts have studied particular classes of Hamiltonians containing long-ranged interactions that exhibit delocalization criticality in lower dimensions\cite{mirlinpre1996,levitov,meltPRL,mobility,izrailevprl1999}. The power-law random-banded matrix (PRBM) ensemble, introduced by Mirlin $\emph{et al.}$\cite{mirlinpre1996} is a large $N\times N$ random
matrix whose entries $H_{ij}$ decrease in a power-law fashion $H_{ij} \propto a_{ij}\left|i - j\right|^{-\alpha}$, with a random hopping coupling constant $a_{ij}$. Mapping the problem onto a nonlinear $\sigma$-model with a non-local interaction, they found a transition from localized to extended states at $\alpha = 1$. Levitov, using a renormalization group approach studied a similar system with disordered site energies and non-random hopping, also found that for  $\alpha > 1$, all states are localized\cite{levitov}, with $\alpha = 1$ as the critical value.

In this article, we  adopt a perturbative approach that is similar in spirit to the original arguments of Anderson, in that a breakdown of perturbation theory in the weak disorder limit indicates that the ground state of the disordered system is very different from the ideal crystal and the two ground-states cannot be approached by adiabatically turning the disorder strength on or off. We modify a scaling argument originally discussed in the reference\cite{domAJP03} for short-ranged systems.
We consider the weak disorder regime and the convergence criteria that determine whether the disorder is allowed to be a weak perturbative effect in the thermodynamic limit is approached but not taken explicitly. If this criterion is rendered false as $N$ increases then all of the eigenstates are localized. We also provide an estimate of the scaling behavior of localization length as function of the disorder strength in the regime where the disorder is moderately strong and nonperturbative.

\section{The non-disordered dispersion}

Consider the tight-binding model with one orbital per site, for $N$ sites with lattice spacing $a$, that is governed by the following Schr${\ddot o}$dinger equation,
\begin{equation}
\varepsilon _l \psi _l  - \sum\limits_z {t_z \psi _{l + z} }  = E\psi _l 
\label{eq:motionideal}
\end{equation}
where $ \psi _l $ are the electron amplitudes at integer site values $l$ and the summation is taken over all unique pairs of lattice sites $z$. The effective width of the spectrum $t_z$ has the power law spatial dependence,
\begin{equation}
t_z  = \frac{{J }}{{\left| z \right|^\alpha  }}
\label{eq:hopping}
\end{equation}
where $J$ is a constant that depends on the details of the atomic orbital overlap matrix elements and the hopping exponent is a positive integer $\alpha = 1, 2, 3, \dots$. It follows that for translationally invariant systems with periodic boundary conditions 
we can make use of the Bloch theorem, yielding a solution to Eq.(\ref{eq:motionideal}) of the form
\begin{equation}
\psi^0 _l (K) = \frac{1}{N}\exp (i{\bf K \cdot R_l} )
\label{eq:bloch}
\end{equation}
where we have introduced the reciprocal lattice vector $ {\bf K} = 2\pi k/Na$, in terms of the wave vector $k$ and ${\bf R_l}=la$. Next, by substituting Eqs.(\ref{eq:hopping}) and (\ref{eq:bloch}) into Eq. (\ref{eq:motionideal}), we have the following expression for the band dispersion:
\begin{equation}
E_{{\bf K},\alpha} ^0   = \left\langle {\varepsilon _0 } \right\rangle  - \frac{{J }}{N}\sum\limits_{n = 1}^\infty  {\sum\limits_{z =  \pm n} {\frac{{e^{\frac{{2\pi i kz}}{{N}}} }}{{\left| z \right|^\alpha  }}} } 
\label{disp1}
\end{equation}
for a mean site energy value, $\left\langle {\varepsilon _0 } \right\rangle$. Eq.(\ref{disp1}) can be evaluated explicitly in terms of polylogarithms as shown in the reference\cite{1d_wc}. Using the definition:
\begin{equation}
Li_n (z) = \sum\limits_{k = 1}^\infty  {\frac{{z^k }}{{k^n }}} 
\label{eq:polylog}
\end{equation}
we arrive at a more compact expression, 
\begin{equation}
E_{{\bf K},\alpha} ^0  = \left\langle {\varepsilon _0 } \right\rangle - J (Li_{\alpha}[e^{i {\bf K}} ] + Li_{\alpha}[e^{ -i {\bf K}} ])  
\label{eq:dispideal}
\end{equation}
It important to note that the fundamental limitation of this approach is that one cannot apply the identity (\ref{eq:polylog}) in the case fractional values of $\alpha$. Therefore, for the entirety of this paper we only consider integer values of $\alpha$ and we concede that a critical value alpha might be a non-integer value however, we can estimate ranges rounded to the nearest integer.

For specific cases of $\alpha$ one can make use of the following exactly summable series\cite{stegun},

\begin{eqnarray}
 \alpha  &=& 1,\rm{}\sum\limits_{n = 1}^\infty  {\frac{{\cos (n\theta)}}{n}}  =  - \ln \left( {2\sin \left( {\frac{\theta}{2}} \right)} \right) \\ 
 \alpha  &=& 2,\rm{}\sum\limits_{n = 1}^\infty  {\frac{{\cos (n\theta)}}{{n^2 }}}  = \frac{{\pi ^2 }}{6} - \frac{{\pi \theta}}{2} + \frac{{\theta^2 }}{4} \\ 
 \alpha  &=& 4,\rm{}\sum\limits_{n = 1}^\infty  {\frac{{\cos (n\theta)}}{{n^4 }}} \nonumber \\ 
&=& \frac{{\pi ^4 }}{{90}} - \frac{{\pi ^2 \theta^2 }}{{12}} + \frac{{\pi \theta^3 }}{{12}} - \frac{{\theta^4 }}{{48}}  
 \end{eqnarray}
This results in the following dispersion relations:

\begin{eqnarray}
 &&E_{{\bf K},1}  = \left\langle {\varepsilon _0 } \right\rangle  - 2J\ln \left( {2\sin \left( {\frac{{\bf K}}{2}} \right)} \right) \\ 
 &&E_{{\bf K},2}  = \left\langle {\varepsilon _0 } \right\rangle  - J\left( {\frac{{\pi ^2 }}{3} - \pi{\bf K} + \frac{{{\bf K}^2 }}{2}} \right) \\ 
 &&E_{{\bf K},4}  = \left\langle {\varepsilon _0 } \right\rangle  - J\left( {\frac{{\pi ^4 }}{{45}} - \frac{{\pi ^2 {\bf K}^2 }}{{6}} + \frac{{\pi {\bf K}^3 }}{{6}} - \frac{{{\bf K}^4 }}{{24}}} \right)  \nonumber \\
 \end{eqnarray}

\section{Perturbative convergence in the presence of disorder}
\subsection{The model}
Having established the dispersion relations for the ideal system, let us now turn our attention the disordered system, where the site energies take random values from an uncorrelated Gaussian distribution. We shall follow precisely, the formalism and notation of F. Dominguez-Adame and V.A. Malyshev in the reference\cite{domAJP03}. The equations of motion result from the following modification to Eq.(\ref{eq:motionideal}),

\begin{equation}
D_l \psi _l  + \left\langle {\varepsilon _0 } \right\rangle \psi _l  - \sum\limits_z {t_z \psi _{l + z} }  = E\psi _l 
\label{eq:dismotion}
\end{equation}
where we have defined $D_l \equiv \varepsilon_l - \left\langle {\varepsilon _0 } \right\rangle$ as the random deviation of the site energies from the mean $\left\langle {\varepsilon _0 } \right\rangle$. $D_l$ is a random variable, distributed according to a standard white-noise, Gaussian distribution:
\begin{equation}
{\cal P}(D_l ) = (2\pi \sigma ^2 )^{ - 1/2} \exp ( - D_l^2 /2\sigma ^2 )
\end{equation}
that is centered at zero $\left\langle D_l \right\rangle = 0$, with a standard deviation $\sigma$. 

\subsection{Scaling the perturbative limit}
The scaling procedure requires two major components. First we must determine the energy level spacing of unperturbed, ideal system, which is dominated by the long wavelength behavior of the dispersion relation, 
$\mathop {\lim }\limits_{K \to 0} E_{K,\alpha }^0 $, that we term $\delta E_{\alpha}^0 (N)$. This quantity solely depends on the mathematical form of the dispersion relations, which ultimately depend on the range of the interactions. Second, we must compute the statistical fluctuations of the disorder-induced energy shifts at the first order in perturbation theory, while assuming that any divergence will occur at the lowest order.

We begin by determining the fluctuations of the first order energy shifts. We will examine the diagonal and off-diagonal contributions separately. Since the site energies still reside on an underlying lattice we can transform Eq.(\ref{eq:dismotion}) to the K-representation by multiplying both sides by (\ref{eq:bloch}) and summing over the site numbers, $l$:
\begin{equation}
(E - E_{{\bf K},\alpha }^0 )\psi ({\bf K}) = \sum\limits_{{\bf K'}} {V_{{\bf KK'}} \psi ({\bf K'})} 
\end{equation}
where we have defined the following,

\begin{eqnarray}
 \psi ({\bf K}) &=& \frac{1}{{\sqrt N }}\sum\limits_l {\psi _l e^{i{\bf K}la} }  \\ 
V_{\bf KK'}  &=& \frac{1}{N}\sum\limits_{l = 0}^{N - 1} {D_l e^{i({\bf K} - {\bf K'})la} }
\label{eq:defv} 
 \end{eqnarray}
Notice that $ V_{{\bf KK'}}$ are simply the matrix elements, of the random potential, or site energies in the unperturbed basis. The diagonal elements $ V_{{\bf KK}}$ represent a first order perturbative energy shift of the ${\bf K}th$ state,  
\begin{equation}
\Delta E_{\bf K}  = V_{\bf KK}  = \frac{1}{N}\sum\limits_{l = 0}^{N - 1} {D_l } 
\end{equation}
which solely depends on the particular quenched realization of the disorder. The square-root fluctuations are given by,
\begin{eqnarray}
 \sqrt {\left\langle {(\Delta E_{\bf K} )^2 } \right\rangle }  &=& \sqrt {\frac{1}{{N^2 }}\sum\limits_{l,l' = 0}^{N - 1} {\left\langle {D_l D_{l'} } \right\rangle } } \nonumber \\ 
  &=& \sqrt {\frac{1}{{N^2 }}\sum\limits_{l = 0}^{N - 1} {\left\langle {D_l ^2 } \right\rangle } } \nonumber \\ 
  &=& \frac{\sigma }{{\sqrt N }}
\label{eq:disscale} 
\end{eqnarray}
Evidently, the diagonal contributions result in an inhomogeneous broadening of the quasi-particle energy level, for given ensemble of quenched disorder.
\begin{figure}
\centerline{\includegraphics[height=2.5in]{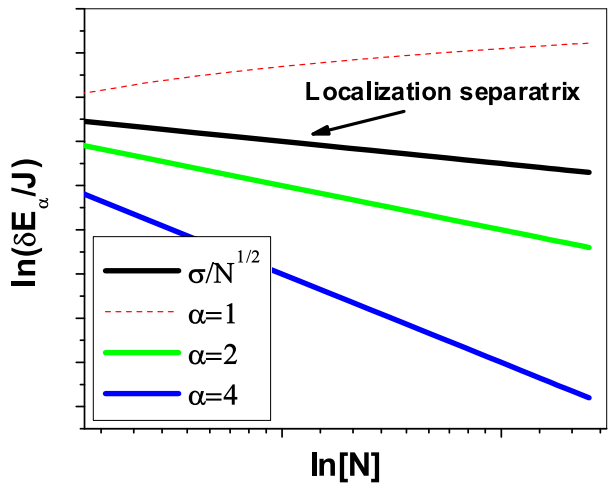}}
\caption{A comparison of the scale dependence of the energy level spacings with the exchange-narrowed fluctuations of the disordered site-energies. Evidently, at $\alpha =1$ the Bloch states of the ideal system satisfy the perturbative condition, increasing faster with increasing $N$ than the critical separatrix. }
\label{fig:pert}
\end{figure}

On the other hand, the off-diagonal components are generally associated with overlapping, or inter-site scattering of the quasi-particle states. Thus, playing an important role in localization. By utilizing Eq.(\ref{eq:defv}), one can see that
\begin{equation}
\sqrt {\left\langle {\left| {V_{\bf KK'} } \right|^2 } \right\rangle }  = \frac{\sigma }{{\sqrt N }}
\end{equation}
In addition, this represents the typical fluctuation, for a given lattice size. Notice that these fluctuations vanish in the limit of a large system size $N \rightarrow \infty$. This effect is known as ``self-averaging" or ``exchange-narrowing". 

Most importantly, our test for localization is the perturbative condition below:
\begin{equation}
\delta E_{{\bf K},\alpha }^0 (N) \gg \sqrt {\left\langle {\left| {V_{\bf KK'} } \right|^2 } \right\rangle }  = \frac{\sigma }{{\sqrt N }}
\label{eq:condition}
\end{equation}
The left hand side of side of Eq.(\ref{eq:condition}) has the following scale dependence for a given value of $\alpha$,
\begin{eqnarray}
 \delta E_{{\bf K},1}^0  &\simeq&  - 2J\ln \left( {\bf K} \right) \propto J\ln \left( N \right) \\ 
 \delta E_{{\bf K},2}^0  &\simeq& 2J\pi {\bf K} \propto \frac{J}{N} \\ 
 \delta E_{{\bf K},4}^0  &\simeq& \frac{{J\pi ^2 {\bf K}^2 }}{6} \propto \frac{J}{{N^2 }}  
 \end{eqnarray}
Let us now examine the energy level spacings and if they satisfy the condition (\ref{eq:condition}), for increasing system size $N$,

\begin{eqnarray}
 \mathop {\lim }\limits_{N \to \infty } &\delta E_{{\bf K},1}^0&  \gg \frac{\sigma }{{\sqrt N }} \\ 
 \mathop {\lim }\limits_{N \to \infty } &\delta E_{{\bf K},2}^0&  < \frac{\sigma }{{\sqrt N }} \\ 
 \mathop {\lim }\limits_{N \to \infty } &\delta E_{{\bf K},4}^0&  < \frac{\sigma }{{\sqrt N }}  
 \end{eqnarray}
Apparently, for $\alpha = 1$ the perturbative condition is satisfied with increasing $N$, therefore the mixing of the Bloch states is weak enough such that the quasi-particle states are expected to remain extended over the entire system length. However, for $\alpha = 2,4$ the disorder has a nonperturbative effect and the perturbative condition (\ref{eq:condition}) is violated such that the Bloch states are strongly mixed and the quantity $\frac{\sigma }{{\sqrt N }}$ decreases slower than $\delta E_{{\bf K},2}^0$ and $\delta E_{{\bf K},4}^0$ signifying that the disordered eigenfunctions cannot be perturbed from the extended Bloch states and thus, all of the eigenstates are localized. A useful illustration is shown in Fig.\ref{fig:pert}, where the scale dependence of the energy level spacings for each $\alpha$ case is plotted on a logarithmic scale. Clearly, as the range of the interactions is decreased with increasing $\alpha$, the eigenstates are more strongly localized.

\section{Scaling at the onset of localization}

\subsection{The localization length}

The threshold at which the condition (\ref{eq:condition}) fails indicates the onset of localization, where localization length is smaller than the size of the lattice. This clearly occurs in the non-perturbative regime, where strength of the disorder is strong but finite and intermediate between the weak and strong disorder regimes. Therefore, one can estimate how the localization length scales as a function of the ratio $J /\sigma$, by solving for $N=\xi$ in the equation below,
\begin{equation}
\delta E_{{\bf K},\alpha }^0 (\xi) = \frac{\sigma }{{\sqrt \xi }}
\label{eq:condition2}
\end{equation}
The solution to Eq.(\ref{eq:condition2}), can be understood graphically in Fig.\ref{fig:pert} as the intersection of each of the $\alpha = 1,2,4$ curves with the separatrix. 

For the $\alpha=1$ case, the localization length solves the equation below,
\begin{equation}
\frac{\xi }{{2\pi }} = \exp \left[ {\frac{\sigma }{{2J\sqrt \xi  }}} \right]
\label{eq:exp}
\end{equation}
It follows that for sufficiently weak disorder, the ratio $\sigma /J \ll 1$, therefore Eq.(\ref{eq:exp}) reduces to the cubic equation, 
\begin{equation}
\frac{{\xi ^3 }}{{(2\pi )^2 }} - \frac{{\xi ^2 }}{{(2\pi )^2 }} + \xi  - \left( {\frac{\sigma }{{2J}}} \right)^2  = 0
\end{equation}
which, admits nonreal, complex solutions and as expected the localization length should not scale with $\sigma /J$ given that all the states are extended. On the other hand, in the opposite limit, $\sigma /J \gg 1$, the solution involves an essential singularity, also admitting complex solutions, therefore the proper scaling of the localization occurs in the highly nonperturbative regime, where our Eq.(\ref{eq:exp}) must be determined numerically. This breakdown of the localization length scaling in both asymptotic limits suggests there is a phase transition connecting the strong and weak disorder regimes.

Next for the strictly localized cases $\alpha =2,4$, for shorter ranged hopping, the scaling of the localization length becomes,
\begin{eqnarray}
 \xi _{\alpha  = 2}  \propto \left( {4\pi } \right)^2 \left( {\frac{J}{\sigma }} \right)^2  \\ 
 \xi _{\alpha  = 4}  \propto \left( {\frac{{2\pi ^4 }}{3}\frac{J}{\sigma }} \right)^{2/3}   
 \end{eqnarray}
As expected when $ \xi _{\alpha  = 4}$, which is more abrupt on the scale of the lattice, one recovers the scaling exponent of $2/3$ which is consistent with numerical simulations of disordered 1D chains\cite{domAJP03} with nearest neighbor hopping. It should be noted that the numerical prefactors depend on the specifics of the model and the proper choice of boundary conditions.

\section{Conclusion}

Lastly, we mention that the methods outlined in the paper can also be generalized to incorporate correlated disorder, for instance, with power law decay. Instead of a white noise distribution of the random site energies, consider a scenario where the power spectrum of the noise has $S(k)\propto 1/k^\beta$, recently discussed as possibly occurring in conducting DNA wires and in conducting 1D systems\cite{nature}. Evidently, the autocorrelation function of the matrix elements of $V_{\bf KK'}$ as designated by Eq.(\ref{eq:defv}) would not scale as $1/\sqrt{N}$, rather the fluctuations of the disorder would decrease faster, depending on the value of $\beta$. This implies that for longer ranged correlated site energies, the eigenstates can more easily satisfy the perturbative condition given by Eq.(\ref{eq:condition}), and thus longer ranged correlations correspond to weaker localization, ultimately allowing for critical delocalization at some particular value of $\beta$.  

To conclude we have demonstrated a more physically transparent approach without the use of non-linear $\sigma$ models or renormalization group methods for analyzing the localization criteria for disordered tight-binding Hamiltonians with power-law hopping. To construct our perturbative approach we relied upon exactly summable series for the integer values of the hopping exponent $\alpha$, allowing for a precise determination of the unperturbed dispersion relations. Then we utilized a simple scaling inequality, which compares the fluctuations of the disorder to the energy level spacings of the unperturbed system. We have found that for $\alpha=1$, there are extended states and the perturbative condition holds for sufficiently weak disorder.

\begin{acknowledgments}

Thanks to Prof. Joseph Rudnick, for useful discussions and assistance. This work was supported by UC General Funds: 4-404024-RJ-19933-02

\end{acknowledgments}

\end{document}